\documentclass{PoS}
\usepackage{amsmath}

\newcommand {\beq} {\begin{equation}}
\newcommand {\eeq} {\end{equation}}
\newcommand {\bea}{\begin{eqnarray}}
\newcommand {\eea}{\end{eqnarray}}

\newcommand {\Tr}{\mbox{Tr\,}}

\newcommand {\ov}{{\rm ov}}

\newcommand{\cN}{{\cal N}}

\title{
\begin{picture}(0,0)(0,0)%
  \put(265,75){\makebox(0,0)[l]{\textnormal
{\normalsize OU-HET-733-2011, ~KEK-CP-260}}}%
 \end{picture}%
Lattice study of 4d ${\cal N}=1$ super Yang-Mills theory with dynamical overlap gluino}

\ShortTitle{Lattice study of 4d ${\cal N}=1$ SYM with dynamical overlap gluino}

\author{
   JLQCD Collaboration:
   \speaker{S.-W.~Kim}$^{a}$\thanks{E-mail: sang@het.phys.sci.osaka-u.ac.jp},
   H.~Fukaya$^a$,
   S.~Hashimoto$^{b,c}$,
   H.~Matsufuru$^{b}$
   J.~Nishimura$^{b,c}$
   and
   T.~Onogi$^a$
   \\
   \\
   \\
   \llap{$^a$}
   Department of Physics, Osaka University,
   Toyonaka, Osaka 560-0043 Japan
   \\
   \llap{$^b$}
   High Energy Accelerator Research Organization (KEK),
   Tsukuba 305-0801, Japan
   \\
   \llap{$^c$}
   School of High Energy Accelerator Science,
   The Graduate University for Advanced Studies (Sokendai),
   Tsukuba 305-0801, Japan
}

\abstract{
We report on a lattice simulation result for four-dimensional ${\cal N}=1$
SU(2) super Yang-Mills theory with the dynamical overlap gluino.
We study the spectrum of the overlap Dirac operator at three different gluino
masses $m=0.2,\,0.1$ and $0.05$ with the Iwasaki action on a $8^3 \times 16$ lattice.
We find that the lowest eigenvalue distributions are in good agreement with
the prediction from the random matrix theory. 
Moreover the mass dependence of the condensate is almost constant,
which gives a clean chiral limit.
Our results for the gluino condensate in the chiral limit
is $\langle \bar \psi \psi \rangle r_0^3 = 0.63(12)$, where $r_0$ is the Sommer scale.
}

\FullConference{ The XXIX International Symposium on Lattice Field Theory - Lattice 2011\\
July 10-16, 2011\\
Squaw Valley, Lake Tahoe, California}

\begin{document}

\section{Introduction}
The $\cN=1$ global supersymmetry in four dimension is
phenomenologically interesting \cite{Witten:1981nf}.
The $\cN=1$ super Yang-Mills theory (SYM) is a simple model
containing only gluon and gluino.
The model is nontrivial since it is expected to have interesting nonperturbative properties
like confinement and discrete chiral symmetry breaking.
An important physical quantity is the gluino condensate
which is an order parameter for the spontaneous breaking
of the discrete chiral symmetry.
If we extend the model to include matters, the Witten index \cite{Witten:1982df}
can vanish implying that a dynamical supersymmetry breaking can occur
with the condensate as an order parameter \cite{Nelson:1993nf}.

Lattice gauge theory could be a means to study the nonperturbative
aspects of supersymmetry.
However the lattice discretization explicitly breaks supersymmetry
and the supersymmetry violating operators at the quantum level
give rise to the fine tuning problem in general.
The $\cN=1$ SYM is exceptional,
since the mass term is the only relevant supersymmetry violating operator,
which can be controlled by taking the chiral limit \cite{Curci:1986sm}.
The model has been studied on the lattice with the Wilson fermion
\cite{Montvay:1996pz}
and the domain wall fermion
\cite{Fleming:2000fa} so far.
Still these approaches have difficulties in obtaining precise chiral limit
due to an explicit breaking of chiral symmetry or a large residual mass.
Even on the lattice, the question whether the R-symmetry of the model
is spontaneously broken or not, remains unclear.

Through this work we provide a numerical evidence for nonzero
gluino condensate in $\cN=1$ SYM by a nonperturbative lattice study.
The overlap fermion \cite{Neuberger:1997fp} is a natural choice for the model
because of its exact lattice chiral symmetry \cite{Ginsparg:1981bj}.
In this study, we use the same strategy as our previous studies in QCD \cite{Fukaya:2009fh},
where clean chiral extrapolation for the chiral condensate
was achieved by dynamical simulation of the overlap fermion
and comparing the Dirac spectrum with the Banks-Casher relation \cite{Banks:1979yr}
and the random matrix theory (RMT) \cite{Verbaarschot:1994qf}.
The key benefit of this strategy is that it is free from both of
the UV and IR problems
when we determine the physical value of the chiral condensate,
as we describe later.

\section{The model and lattice formulation}
The Euclidean action for $\cN=1$ SYM is
\bea
S=\int d^4 x \, \Tr \Big \{ \frac{1}{4} (F_{\mu\nu})^2
+ \frac{1}{2} \bar\psi \gamma^\mu D_\mu \psi \Big \}~.
\label{action}
\eea
Here $\psi = \psi^a (x) \, T^a$ denotes the gluino field,
which is an adjoint Majorana fermion satisfying $\bar\psi=\psi^T C$.
The action is invariant under the supersymmetry transformation
\bea
\delta A_\mu = -2 g \bar\psi \gamma_\mu \epsilon ~,~~
\delta \psi = -\frac{i}{g} \sigma^{\mu\nu} F_{\mu\nu} \epsilon ~,
\label{susy}
\eea
and the R-symmetry
\bea
\psi \rightarrow e^{-i \alpha \gamma_5} \psi ~,~~
\bar\psi \rightarrow \bar\psi e^{-i \alpha \gamma_5} ~,
\label{Rsymmetry}
\eea
which coincides with the $U(1)$ axial symmetry.
This $U(1)_A$ is anomalous and the axial current
$J_\mu = \bar\psi \gamma_\mu \gamma_5 \psi$ satisfies
\bea
\partial^\mu J_\mu = \frac{N_c g^2}{16 \pi^2} F^{a\mu\nu} \widetilde F^a_{\mu\nu} ~.
\label{anomaly}
\eea
The anomaly leaves discrete subgroup $Z_{2N_c}$ invariant at quantum level
but it can be spontaneously broken down to $Z_2$.
An order parameter is the gluino condensate $\langle \bar\psi \psi \rangle$.
No Nambu-Goldstone boson appears since $Z_{2N_c}$ is a discrete group.

On the lattice the overlap Dirac operator has no sign problem,
since it possesses Ginsparg-Wilson relation,
$\gamma_5$ hermiticity, and two-fold degeneracy from the adjoint representation.
Therefore we can write the fermionic action as
\bea
&&S_f = -\frac{1}{4} \log \, \det (D_\ov^\dag D_\ov) ~,
\label{actionF}
\eea
where the overlap Dirac operator $D_\ov$ in the adjoint representation
is constructed from the Wilson Dirac operator $D_W$ as
\bea
&&D_\ov = M_0 \Big ( 1+ D_W \big / \sqrt{D^\dag_W D_W} ~ \Big ) ~,
\label{overlap}\\
&&(D_W)^{ba}_{yx} = \delta_{yx}\delta^{ba} -K \sum_{\mu=1}^4
[\delta_{y,x+\hat\mu} V^{ba}_{x\mu} + \delta_{y+\hat\mu,x} (V^{ba}_{y\mu})^T] ~.
\label{wilson}
\eea
$V^{ba}_{x\mu}$ is the adjoint link variable defined as
$V^{ba}_{x\mu} = 2\Tr (U^\dag_{x\mu}T^b U_{x\mu} T^a)$
and the hopping parameter is set to $K = \frac{1}{8-2M_0}$.
We introduce an explicit mass $m$ for gluino in order for a comparison
 with the random matrix models.
\section{Banks-Casher relation and random matrix theory}
The Banks-Casher relation \cite{Banks:1979yr} relates
 the Dirac spectrum to the chiral condensate.
Moreover the details of low-lying Dirac eigenvalue distributions are
described by the chiral random matrix theory (ChRMT) \cite{Verbaarschot:1994qf}
including
finite volume corrections to the Banks-Casher formula.

There are two advantages in extracting the condensate via ChRMT.
One is that the Dirac spectrum is free from UV power divergences,
which we can explicitly see in the extended Banks-Casher formula :
\bea
%
%
&&\pi \rho (\lambda)
=  \lim_{\epsilon \rightarrow 0}  \tfrac{1}{2} \big (
- \langle \bar \psi \psi \rangle |_{m=i\lambda+\epsilon}
+ \langle \bar \psi \psi \rangle |_{m=i\lambda-\epsilon} \big )~.
\label{BanksCasher}
\eea
Another advantage is that the finite volume effects and finite mass effects
can be corrected by ChRMT.
In RMT, the low-lying eigenvalue distributions are
characterized by the dimensionless parameter $\mu=m \Sigma V$.
For any given volume $V$ and mass $m$,
the condensate $\Sigma$ is the only free parameter
that is defined in the infinite volume and massless limits.

The gluino is an adjoint representation fermion and the corresponding Dirac operator
satisfies $[D,C^{-1}K]=0$, where $C$ is the charge conjugation and $K$ is the complex
conjugation operator. Consequently the universality class for $\cN=1$ SYM
 is different from QCD, and it belongs to the chiral symplectic ensemble (ChGSE).
The partition function is the same as that for chiral unitary ensemble, which describes QCD,
except that the matrix elements are real quaternion instead of complex number.
This is equivalently described by partially quenched chiral perturbation theory.

For later convenience, we present the first eigenvalue distribution of ChGSE
for zero topological charge, 
\bea
&&p_1 (\zeta,\mu) = c \, e^{-\zeta^2/2} \, \zeta \, \tilde \mu^4
\frac{\int^1_0 dt \, \tilde \mu^{-3} \, I_3 (2t\tilde \mu)}
{\int^1_0 dt \, \mu \, I_{-1} (2t \mu)} ~,
\label{lowest_distribution}
\eea
where $\zeta = \lambda \Sigma V$, $\mu = m \Sigma V$,
$\tilde \mu = \sqrt{\zeta^2+\mu^2}$, $c$ is a normalization constant,
and $\lambda$ is the eigenvalue.
The general formula for the higher eigenvalues is given in \cite{Damgaard:2000ah}.
We note that the quenched limit can be obtained by simply taking
the limit $\mu \rightarrow \infty$.

\section{Results for Dirac spectrum and gluino condensate}
Our dynamical simulation for $\cN=1$ SYM
with the overlap fermion is performed on a $8^3 \times 16$ lattice
with $SU(2)$ gauge link fields.
For this small lattice, we need to carefully pick up a gauge action
and the parameters $\beta, \, M_0$ to avoid the Aoki phase.
The $\beta$ is chosen in such a way that the lattice
has a sufficiently large physical  volume,
and the $M_0$ is chosen to guarantee the locality.
The hermitian Wilson-Dirac spectrum for the quenched theory
suggests that the region for an implementation of the
overlap Dirac operator inside two fingers of the Aoki phase is
narrower in the present case of the adjoint representation
than in the case of the fundamental representation.
Based on a Dirac spectrum survey in the quenched theory,
we use the Iwasaki gauge action 
with $\beta=1.05$ and $M_0=1.9$.

The overlap fermion has exact zero modes which are related to
the topological charge of the gauge fields by the Atiyah-Singer index theorem.
The topology plays a significant role on the small volume for a comparison to ChRMT.
In our simulation, the global topological charge is fixed
by introducing extra Wilson fermions to produce a Boltzman weight
$\det \frac{H_W^2}{H_W^2 +m_{tm}^2}$
where $H_W=\gamma_5 D_W$ is the hermitian Wilson-Dirac operator
with $K = \frac{1}{8-2M_0}$.
This term suppresses the near zero modes of $H_W$ and generates a gap,
which supports the exponential locality of the overlap Dirac operator
and reduces the numerical cost for the sign function in the overlap Dirac operator
\cite{Fukaya:2006vs}.
The inclusion of $\det \frac{H_W^2}{H_W^2 +m_{tm}^2}$ should not affect
the low energy physics.
We select zero topological charge and small twisted mass of $m_{tm}=0.01$.
The choice of the twisted mass does not affect the low-lying Dirac spectrum in the quenched theory.

\begin{table}[tb]
  \centering
  \begin{tabular}{c|ccccccc}
    \hline
    $m$ & $d \tau \times N_\tau$ & $N_{\rm trj}$ & $accept$ &
    $\langle e^{-\Delta H} \rangle$ & $\langle P \rangle$  & $r_0/a$  \\    \hline
    0.20 & 0.25$\times$4 & 396 & 0.700(29) & 0.977(58) & 0.68747(21) & 4.34(17)  \\
    0.10 & 0.20$\times$5 & 372 & 0.850(15) & 0.971(18) & 0.68752(19)  & 5.25(30)  \\
    0.05 & 0.25$\times$4 & 382 & 0.658(26) & 0.996(42) & 0.68795(16)  & 5.44(33)  \\
    \hline
  \end{tabular}
  \caption{Summary of the simulation. 
$m$ is the gluino mass,
$N_{\rm trj}$ is the number of thermalized trajectory,
$P$ is the plaquette, and $r_0$ is the Sommer scale.}
  \label{table_ensemble}
\end{table}

With the Hybrid Monte Carlo algorithm,
we generated ensembles with three different gluino mass 
as shown in the table \ref{table_ensemble}.
For the low-lying eigenvalue distribution of the overlap Dirac operator,
28 lowest eigenvalues are measured at every other trajectory
by implicitly restarted Lanczos method.
Note that the adjoint representation gives two-fold degeneracy,
and the $\gamma_5$ hermiticity enforces eigenvalues to form
complex conjugate pairs. Therefore there are essentially only
7 different eigenvalues among 28.
Our main interest is in the lowest eigenvalue,
for which the data shows clear complex conjugate pairing with two-fold degeneracy.

The integrated autocorrelation time is estimated
by the history of the lowest Dirac eigenvalue,
which is smaller than 20 trajectories for all three input gluino masses.
All statistical errors are estimated by the jackknife method
with a binsize of 20 trajectories.

\begin{figure}[tb]
\begin{center}
\includegraphics[height=4.8cm]{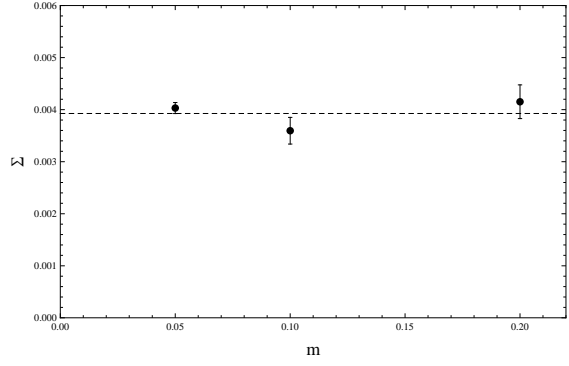}
\end{center}
\caption{The condensate determined by (4.1)
is plotted for each gluino mass.
The dashed line shows the constant fit, which is in good agreement with the data points.}
\label{LSYMcondensate}
\end{figure}

The relationship between ChRMT and the chiral perturbation theory (ChPT)
suggests an identification
$\zeta=\lambda \Sigma V$, from which we can extract the condensate.
The condensate at each gluino mass is determined by comparing the average
eigenvalues of lattice data and that of ChGSE.
\bea
&& \langle \zeta \rangle_{\rm RMT}
= \int^\infty_0 d \zeta \, \zeta \, p_1 (\zeta, m \Sigma V)
= \langle \lambda \rangle_{\rm lat} \Sigma V ~.
\label{average_eigenvalue}
\eea
Figure \ref{LSYMcondensate} shows the gluino mass dependence of $\Sigma$.
The comparison with ChGSE gives almost the same values of
the gluino condensate for different input gluino masses.
The mass dependence is weaker than that in QCD,
which can be explained by the absence of Nambu-Goldstone modes,
or the absence of chiral logarithms.
The constant fit gives $\Sigma = 0.00393(18)$.
With the Sommer scale obtained from the lightest mass ($m=0.05$),
we obtain $\langle \bar \psi \psi \rangle r_0^3 = 0.63(12)$
in the chiral limit.
\footnote{The value $\Sigma^{1/3} r_0 = 0.86(5)$ is slightly larger than
that for QCD in our previous study : $\Sigma^{1/3} r_0 \approx 0.6$.}

\begin{figure}[tb]
\begin{center}
\includegraphics[height=4.8cm]{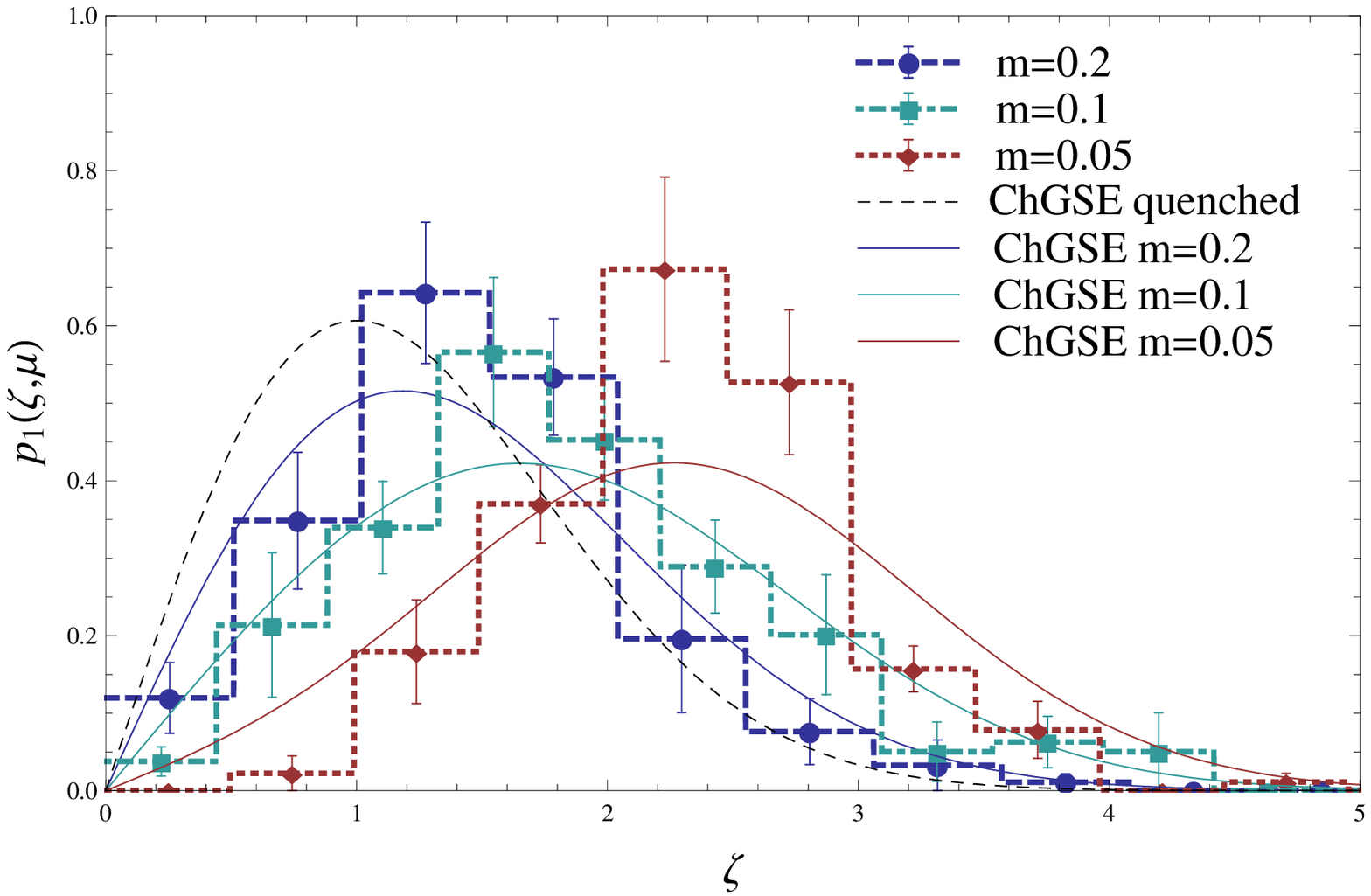}~~~
\includegraphics[height=4.8cm]{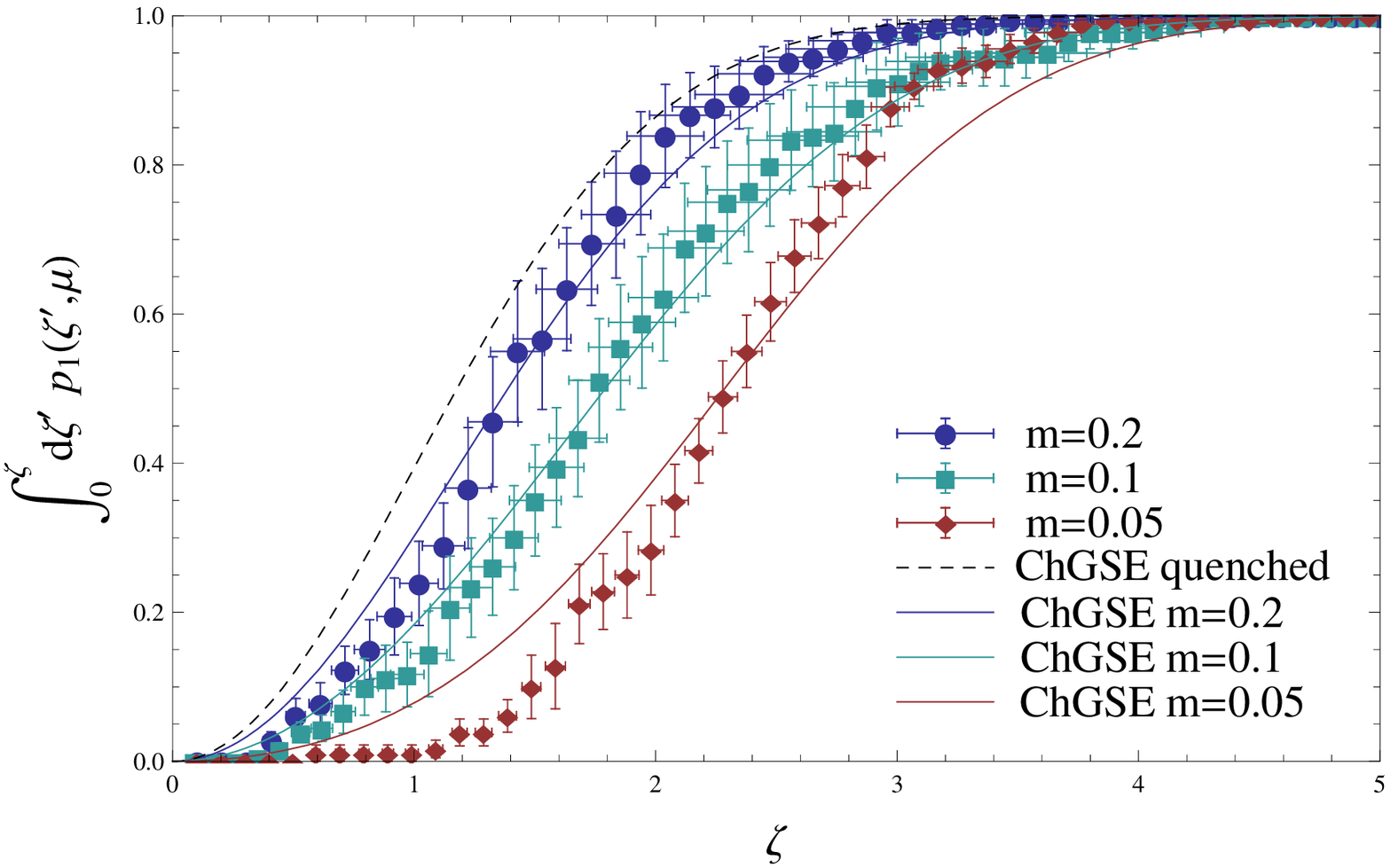}
\end{center}
\caption{The lowest eigenvalue distribution is compared with ChGSE.
Curves are the predictions from the ChGSE for various masses.
The condensate determined from (4.1) for each mass was used as an input.
The horizontal errorbar (right) reflects the uncertainty of the condensate.}
\label{LSYMrmt}
\end{figure}

Let us look into more details of the eigenvalue distribution.
Figure \ref{LSYMrmt} shows the lowest eigenvalue distribution $p_1(\zeta,\mu)$ (left)
and the cumulative distribution $\int_0^\zeta d \zeta' p_1(\zeta',\mu)$ (right)
for each mass, where we used the determined condensate value
as an input ($\mu=m \Sigma V$).
The distributions show clear mass dependence,
and the data at $m=0.1$ and $0.2$ agree well with the prediction from the ChGSE.
There is a common tendency that the distribution is narrower
than the predicted curves,
and this tendency becomes clearest for the lightest mass data.
Since ChGOE or ChGUE has wider distribution compared to ChGSE,
ChGSE fits best to the data anyway.
This deviation is most likely from the higher order corrections of ChPT,
which is not considered in RMT.

\begin{figure}[tb]
\begin{center}
\includegraphics[height=4.8cm]{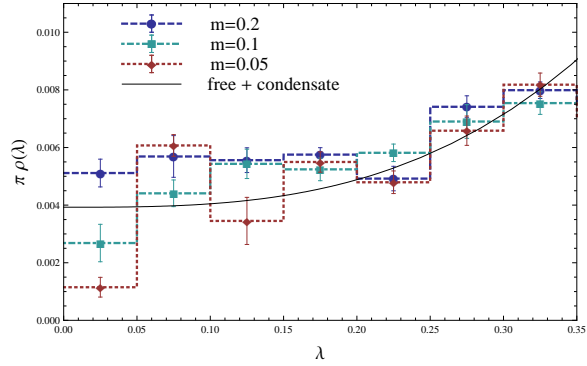}
\end{center}
\caption{The low-lying eigenvalue distribution is shown for each mass.
Black curve shows $\Sigma+\frac{3}{8 \pi^2} \lambda^3$,
where the second term comes from the free theory in 4-dimension.
}
\label{LSYMevdist}
\end{figure}

Figure 3 shows the low-lying eigenvalue distribution,
which can be roughly divided into three regions :
the lowest edge near $\lambda \approx 0.05$, a plateau region between $0.05<\lambda<0.25$,
and an increasing region for $\lambda>0.25$.
The overall shape is consistent with the behavior
$\frac{3}{8 \pi^2} \lambda^3$ of the free theory
plus the condensate $0.00393(18)$, which is shown as the black curve in the figure.
Note that the plateau values, which are without leading volume corrections,
do not depend on masses very much.
Since the difference from these values
to the obtained condensate value in the chiral limit is around $20\%$,
we expect that the systematic error from the higher order effects are less than $20\%$.

\section{Summary}
We have performed a dynamical simulation of 4d $\cN=1$ SYM
with the overlap fermion for the gluino on a $8^3 \times 16$ lattice.
The exact chiral symmetry of the overlap fermion
and the prediction from ChRMT allow us to extract the gluino condensate
$\langle \bar \psi \psi \rangle r_0^3 = 0.63(12)$ in the chiral limit. 
The lowest Dirac eigenvalue distribution for two heavy masses ($m=0.2, 0.1$)
nicely matches with ChGSE.
For the lightest gluino mass ($m=0.05$) data, the distribution shows a deviation from
the prediction possibly due to corrections from the next order finite volume effect.
The expectation value of the lowest Dirac eigenvalue is used to determine the condensate,
and the obtained condensate for each mass does not much depend on the mass.
This confirms that our strategy works very well in the $\cN=1$ SYM,
since there is no Nambu-Goldstone boson nor chiral log effect.
The low-lying eigenvalue distribution is consistent with the simple sum
of the constant (condensate) and the free theory result.

The only relevant supersymmetry violating operator in the model is the mass term,
which can be controlled by taking the chiral limit. 
Therefore it is interesting to see the violation of the supersymmetry Ward identity
and its dependence on the input gluino mass.

We thank P. H. Damgaard, J. Giedt, G. Bergner, Y. Kikukawa, and N. Yamada
for useful discussions.
Numerical simulations are carried out on
IBM Blue Gene/L at KEK under the support of its Large-scale
Simulation Program (Nos. 09/10-09, 10-11) and Hitachi SR16000 at YITP.
This work is supported in part by the Grant-in-Aid from the Japanese Ministry of
Education, Culture, Sports, Science and Technology
(No.\,20105002, 20105005, 21674002, 23105710),
the Grant-in-Aid for Scientific Research (No.\,20540286, 23244057) from JSPS,
and the HPCI Strategic Program of Ministry of Education.


\end{document}